# Prioritising Data Items for Business Analytics: Framework and Application to Human Resources

Tom Pape* a,b

a Department of Mathematics, University College London, 4 Taviton Street, London WC1H 0BT, UK
b Concentra Consulting Ltd., Thames House, 18 Park Street, London SE1 9EQ, UK

**Abstract:** The popularity of business intelligence (BI) systems to support business analytics has tremendously increased in the last decade. The determination of data items that should be stored in the BI system is vital to ensure the success of an organisation's business analytic strategy. Expanding conventional BI systems often leads to high costs of internally generating, cleansing and maintaining new data items whilst the additional data storage costs are in many cases of minor concern – what is a conceptual difference to big data systems. Thus, potential additional insights resulting from a new data item in the BI system need to be balanced with the often high costs of data creation. While the literature acknowledges this decision problem, no model-based approach to inform this decision has hitherto been proposed. The present research describes a prescriptive framework to prioritise data items for business analytics and applies it to human resources. To achieve this goal, the proposed framework captures core business activities in a comprehensive process map and assesses their relative importance and possible data support with multi-criteria decision analysis.

## 1 Introduction

Business functions such as customer service, human resources (HR), IT, legal support, marketing, procurement, production, project delivery, R&D and sales can draw on widely used business intelligence (BI) system solutions. BI systems can be seen as the technological foundation to conduct business analytics (Chiang et al. 2012, Lim et al. 2012). Liberatore and Luo (2010) define business analytics as a set of methods that transform raw data into action by generating insights for organisational decision making. For instance, knowing that younger women are more likely to buy a pricier version of a certain product helps marketers target advertisements appropriately. A widely endorsed representation of business analytics observes organisations mature from using descriptive quantitative data analyses to explain what is happening now through predictive analyses to estimate what will happen in the future to prescriptive analyses using management science tools to help decide what to do next (Delen & Demirkan 2013, Liberatore & Luo 2011, Lustig et al. 2010). All modern BI systems provide efficient data storage, allow data exploration with descriptive statistics using aggregation and grouping commands and present the results on attractive management dashboards. In addition, more advanced BI systems possess ready-to-use predictive data mining tools. Prescriptive analytics often require highly specialised add-ons for a BI system (e.g., pricing and inventory management tools).

Operational research techniques have supported the development of BI systems in multiple ways (Mortenson et al. 2015). Tracking and visualising key performance indicators (KPIs) is a traditional function of BI systems, and operational research has been employed to develop and structure such KPIs (Fortuin 1988, Smith & Goddard 2002). Optimisation methods play an important role in data mining, which in turn is commonly used in BI systems (Corne et al. 2012, Olafsson et al. 2008). 'Soft' OR has been applied to the design of new BI systems (e.g., Martínez et al. 2013, Sørensen et al. 2010), and decision analysis has informed the evaluation of BI systems (e.g., Lin et al. 2009, Rouhani et al. 2012). Finally, operational research methods such as forecasting, scenario analysis and simulation have been implemented on top of BI systems to enhance organisations' business analytic capabilities (e.g., Negash & Gray 2008, Telhada et al. 2013).

---

* Email: t.pape@surrey.ac.uk.

This study addresses the decision problem of data items that a business function should store in its BI system to conduct business analytics. As most BI data is internally created, expanding a BI database leads to high administrative costs for generating, cleansing and maintaining new data items whereas data storage costs are typically of much lower magnitude – a conceptual difference to big data systems. The high costs of data creation require business function to be selective. We propose a prescriptive framework based on multi-criteria decision analysis to help business functions make this selection decision, i.e. to help business functions decide 'what to do next' in their analytic strategy to use the language of the aforementioned classification.

To demonstrate the usefulness of the proposed framework, we apply it to HR. HR analytics are still in their infancy with a heavy focus on descriptive analytics (e.g., Infohrm 2010, Smith 2013), some applications of predictive analytics (e.g., Cascio & Boudreau 2011, Fitz-enz 2010) and little research in prescriptive analytics (e.g., Bordoloi & Matsuo 2001, Canós & Liern 2008). However, the interest in business analytics has grown tremendously in many HR departments within the last few years (Aral et al. 2012, Levenson 2011). The launches of several HR BI systems such as Fusion, OrgVue, SuccessFactor and WorkDay reflect this development.

This paper is organised as follows: Section 2 highlights the practical relevance of a framework to prioritise data items based on a literature review and our experience. Section 3 describes the framework in detail. Section 4 is dedicated to the application of the framework to HR. Section 5 concludes the study.

## 2 Practical relevance[1]

The selection of data items that should be acquired and managed in a BI system to conduct business analytics is a significant challenge (e.g., Ballou & Tayi 1999, Kirsch & Haney 2006, Lawyer & Chowdhury 2004, Loeb et al. 1998, March & Hevner 2007, Mishra et al. 2013, Sahay & Ranjan 2008, Sen & Sen 2005, Watson et al. 2004). Ultimately, the success of any BI system relies on available data (Ramakrishnan et al. 2012). Although countless analyses (or metrics) for business data circulate in professional and academic literature, the problem of raw data that business functions should routinely collect in BI systems is surprisingly underexplored. However, this problem is of high practical relevance for four reasons. First, not all data items are equally useful. Business functions can usually draw on some multi-purpose data items that support many analyses, whereas other data items are only required for one particular analysis. Not surprisingly, multi-purpose data items tend to be judged as more relevant by practitioners in our experience and are therefore more frequently collected (see section 4.3). Second, the extract-transform-load (ETL) process of cleansing data items from legacy systems and external sources to transfer them to a new BI system typically accounts for more than 50% of the time and costs of a BI project (Davenport et al. 2001, Shen et al. 2012, Wang et al. 2012). Third, in a study of small and medium enterprises (SMEs), which have previously setup or are in the process of setting up a BI system, 13 of 20 SMEs mentioned a lack of existing data items for business analytics as a major barrier for succeeding in their analytic strategy (Olszak & Ziemba 2012, see also Marx et al. 2011). This observation is in line with our experience when talking to or working with practitioners. However, the creation of these missing data items is even more time consuming and expensive than cleansing and transferring them from an existing database. Fourth, the continuous update of a BI system with recent data can lead to high reoccurring costs (Choudhury & Sampler 1997, Eppler & Helfert 2004, Even et al. 2007, Gilad & Gilad 1985, Morrison 2015). Examples of such costs include administrator time for monitoring that data entries are up to date, gathering customer opinions, license fees for proprietary market data, license and training costs for software packages for the central collection of data items and staff time for typing details about project milestones. Thus, potentially high costs associated with loading a BI system with data make it imperative that business functions are judicious about data items that they incorporate.

---

[1] This section is informed by a systematic literature search conducted in August 2015 using the Google Scholar functions. The 86 information system journals listed by the Association for Information Systems (2012) were chosen as a source. This list also includes several management and management science journals. The full-text search term (('business analytics' OR 'business intelligence' OR 'data warehouse') AND ('data requirements' OR 'information requirements' OR 'data selection' OR 'data prioritisation' OR 'data prioritization')) was used. All publications that discuss theoretical and practical aspects of the data selection problem for BI systems or address this problem in a case study are considered (inclusion criteria). The search identified 261 publications, of which 189 were excluded based on title and abstract and 54 after a full-text review. The remaining 18 publications that met the inclusion criteria are asterisked in the reference list and cited in section 2 of this paper. Further relevant publications have been identified through an unsystematic literature search using Google Scholar and through reviewing the bibliographies of selected papers.

Drucker (1995) and March and Hevner (2007) noted that managers should be able to generate both internal and external information from BI system's data. They considered internal information to include accounting and financial measures, productivity measures, measures relating to organisation's core competences and competitive advantages and measures relating to the allocation of organisation's scarce resources. On the other hand, they considered external information to include analyses of markets, customers, technologies, worldwide finance data and changes in world economy.

Precise data items required to generate this internal and external information are typically derived from simple surveys, unstructured interviews and qualitative requirement models. Beynon-Davies (2004), Gilad and Gilad (1985), Kim and Gilbertson (2007), Loeb et al. (1998), Loshin (2012) and Shanks and Darke (1999) used surveys and manager interviews in their case studies to learn about users' analytic needs. Qualitative data-requirement models aim at either integrating a selection of existing data sources into the BI system (supply-driven approach), systematically exploring users' data needs (demand-driven approach) or combining both approaches in a meaningful manner. Dori et al. (2008), Romero and Abelló (2010b) and Takecian et al. (2013) developed supply-driven approaches that derive the initial selection of data items from the conceptual model of operational databases. Prakash and Gosain (2008) constructed a demand-driven data-requirement model by breaking down organisation's goals into concrete decisions that are informed by simplified SQL queries (see also Romero & Abelló 2010a). The demand-driven approach from Paim and Castro (2003) employed the Use Case system modelling language to explore how end users are likely to interact with the BI system, which leads to a list of required data items. Giorgini et al. (2008), Maté et al. (2014) and Mazón and Trujillo (2009) applied the widely used $i*$ system modelling language to structure the actions of organisation's stakeholders (as data suppliers) and the goal-oriented decision making of managers (as users in demand of data), which allows gaining an overview of potentially useful data items for the BI system. Maté and Trujillo (2012) and Mazón et al. (2007) also aimed at linking the data supply from existing databases with data requirements expressed by end users by employing advanced modelling languages. All mentioned models are concerned with (i) exploring data items that are potentially useful for the BI system and (ii) the structuring of the data items from a BI system architecture viewpoint. These models do not provide guidance on how the relevance of data items should be systematically assessed.

According to Wetherbe (1991) and in line with the decision analysis and 'soft' OR literature (e.g., Phillips & Phillips 1993, Rosenhead & Mingers 2001), the discussion of data requirements in workshops tends to lead to better outcomes because of a wider knowledge base, a broader perspective and particularly the opportunity for mutual learning about the decision problem. The works of Ormerod (1995, 1996, 1998, 2005) on mixing 'soft' OR techniques to develop information system strategies and of Checkland (1988, Checkland & Holwell 1997, Winter et al. 1995) on the link between soft system methodology and information systems provide good examples of engaging workshop designs; however, these researchers do not explicitly address the problem of selecting data items.

In addition, even workshops as a stand-alone approach, either by allowing end users to create their wish list or by giving them a prepared list of choices, often do not lead to completely satisfying outcomes. When allowing end users to create their wish list of data items, many BI systems initially fail to satisfy managers' expectations because the managers usually do not know in advance what data and analyses they need (Jenkins et al. 1984, Lederer & Prasad 1993, Romero

& Abelló 2010b). The addition of the cleansed and tested missing datasets later to a previously launched BI system often leads to massively higher costs (Wetherbe 1991) and typically takes three months each (Eckerson 2005). Alternatively, when providing managers with a list of choices for potential analyses and required data items, they typically want all of them (Eckerson 2010, Judd et al. 1981).

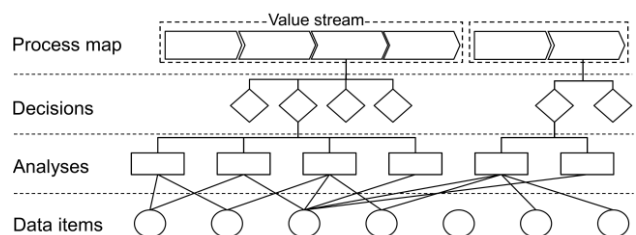

*Fig. 1.* Linking model.

It becomes clear that engaging managers in a discussion about data items that they need in their BI system is required but not sufficient. A decision model which prioritises data items for business analytics would greatly inform such discussions. Our systematic search of the literature indicates that a formal model to prioritise data items for business analytics seems hitherto not to have been published. The present study addresses this gap in the literature.

Finally, it should be noted that the problem described in this study primarily applies to conventional BI systems and not to big data systems.[1] Conventional BI systems are based on the paradigm that the user knows what questions (or hypotheses) they wish to answer and with what kind of analyses. Such BI systems allow the user to conduct their analyses on a well-structured relational database with managed data items using their background knowledge of the business. Thus, it is the user who decides on the relevance of particular data items to answer a particular question—a decision is made *before* accessing the data. Creating new data is often connected with some sorts of costs. In conventional BI systems, the organisation is typically the producer of new data (Chen et al. 2012) and therefore must bear the data creation costs. Whilst the creation of some data items is essential for the organisation's operations (e.g., finance accounts, client address and delivery status), many other data items exist whose creation is optional. Examples include rating of quality of delivered goods (procurement), reasons for investigating expense claims (finance), number and length of client visits (sales), types of problem solutions provided to individual customer calls (customer service) and case mix managed by in-house lawyers (legal support). The primary constraint for amassing large amounts of data in conventional BI systems is the high cost of internal data creation (including ETL).

In contrast to BI systems, big data systems as, for instance, defined by Anderson (2008), Mayer-Schönberger and Cukier (2013) and Katal et al. (2013) employ machine learning on large, diverse and unstructured datasets to make inferences about the probabilities of certain events or associations. In big data systems, it is the machine that decides how relevant particular data items are based on statistical analyses—a decision is made *after* accessing the data. To be effective, big data systems require vast streams of very cheaply generated data items. Sources for this cheap data may include data already generated by the organisation's enterprise resource system, customer interactions with web 1.0 systems (e.g., search terms and clickstreams), social media and RFID chips. The primary constraint for amassing large amounts of data in big data systems is IT costs for storage, computing power and human capital.

## 3 Framework to quantify the value of data items

### 3.1 Linking model

We decompose business analytics into four hierarchical layers: process map, decisions, analyses and data items. The resulting top-down linking model connects data items to processes (see Fig. 1).

---

[1] The distinction between conventional BI systems and big data systems here is analogous to the distinction between BI&A 1.0 and BI&A 2.0/3.0 by Chen et al. (2012). According to a large survey by Kart et al. (2013), only 8% of the organisations have big data solutions in place in some of their business functions.

The structure of the linking model is rooted in three common concepts in the enterprise information system literature. First, models assigning data items to processes can be found in many variants in the literature (e.g., Olson 2003, Scheer et al. 2002) as most data items are primarily used in particular business processes rather than being relevant to all of them. Second, managers usually don't really know what data items they need, as explained in section 2. It is therefore recommended that managers not be asked what data or analyses they want but rather what decisions they have to make (Eckerson 2010, March & Hevner 2007, Wetherbe 1991). Following this advice, important management decisions serve as a bridge between analyses and processes in the top-down linking model. Third, the process/decision-analyses-data hierarchy closely resembles the knowledge-information-data paradigm in enterprise information systems (Tuomi 2000).

BI systems are typically designed to be process oriented to make sure they support all core activates of a business function to a satisfying degree (Golfarelli et al. 2004). Borrowed from supply chain management, the process map divides a business function into value streams, which in turn are further split into single processes. The decision layer captures the core questions to which the business function must be prepared to provide answers. Assigning each decision to a process ensures that the list of decisions considers all relevant business activities. The analysis layer comprises analytical models and metrics, which are of relevance for a particular business function. Each standard decision is informed by a set of analyses. The data-item layer contains the raw numerical and qualitative data required for conducting the analyses. Value-stream-to-process, process-to-decision and decision-to-analysis links are conceptualised as one-to-many relationships, but many-to-many relationships are allowed for analysis-to-data-item links, meaning that an analysis can require multiple data items, and a data item can feed into multiple analyses.

*3.2 Importance weights w*

In most cases, it is not reasonable to assume that organisations pay equal attention to each of the different decisions in the linking model. As an example, consider the following two marketing decisions: 'How should we price our products in a particular market?' and 'How can the effectiveness of our social media copywriting be improved?' Data that feeds the first decision are usually deemed much more valuable than the data required for the second one. This observation is incorporated into the framework by assigning different relative importance weights to decisions. The weighting approach used is explained in detail in two subsections: 'hierarchical weighting' and 'neutral-good swing weighting'.

*3.2.1 Hierarchical weighting*

Comparing the importance of decisions from very different business processes is cognitively demanding and often leads to somewhat arbitrary judgements (Saaty 1990). The finance decisions 'How much cash flow do we need?' versus 'Do we need to intervene on our expense claim policy?' may serve as an example. Therefore, hierarchical weighting is used to break down the challenge so that only decisions from the same process are weighted against each other:

- On the value stream level, the weights $w_h \in [0,1]$ for value streams $h$ are rated against each other such that $\sum_h w_h = 1$.
- On the process level, the weights $w_{hi} \in [0,1]$ for each process $i$ within value stream $h$ are rated against each other. The process weights $w_{hi}$ must total 1 for each value stream, i.e. $\sum_i w_{hi} = 1 \ \forall h$.
- On the decision level, the weights $w_{hij} \in [0,1]$ for each decision $j$ within process $i$ of value stream $h$ are rated against each other. Again, the decision weights $w_{hij}$ must total 1 for each process, i.e. $\sum_j w_{hij} = 1 \ \forall (h, i)$.

This structure follows the well-established value-tree approach in multi-criteria decision analysis (Keeney & Raiffa 1976).

*3.2.2 Neutral-good swing weighting*

Simple direct-importance judgements (e.g., asking participants to rate the priority of each value stream on a scale between 0 [irrelevant] and 10 [absolutely crucial]) regularly lead to distorted results, for two reasons. First, participants might interpret an importance scale quite differently (e.g., what 5 actually means on a 0–10 scale) if not provided with an unambiguous description of the scale's meaning (Belton & Stewart 2002). Second, the value of enhanced decision making in a business activity depends on how important the resulting performance difference is for the organisation rather than its absolute value (Cascio & Boudreau 2011, Keeney 2002). For instance, the HR process 'Administer Payroll' is highly relevant, but the value of improving this HR process by better decision making is very limited for most Western organisations as it is usually already done well enough.

To avoid these two traps, we use swing weighting (Edwards & Barron 1994, von Winterfeldt & Edwards 1986) with neutral-good ranges (Bana e Costa 1996, Bana e Costa et al. 2002). Assume an 'average organisation' of larger size whose capability for making decisions in the business function of interest is only average (median), meaning that 50% of the other organisations are better in making each of the standard decisions and 50% are worse. A decision swing $(h, i, j)$ for decision $j$ in process $i$ of value stream $h$ is defined as the improvement of the average organisation only in its capability of making decision $j$ from 'average' to belonging to the 'top 10%', ceteris paribus.[1] By using swings, all participants base their importance assessments on a sufficiently shared perception of how far an average organisation's decision-making capabilities can be realistically enhanced in each area.

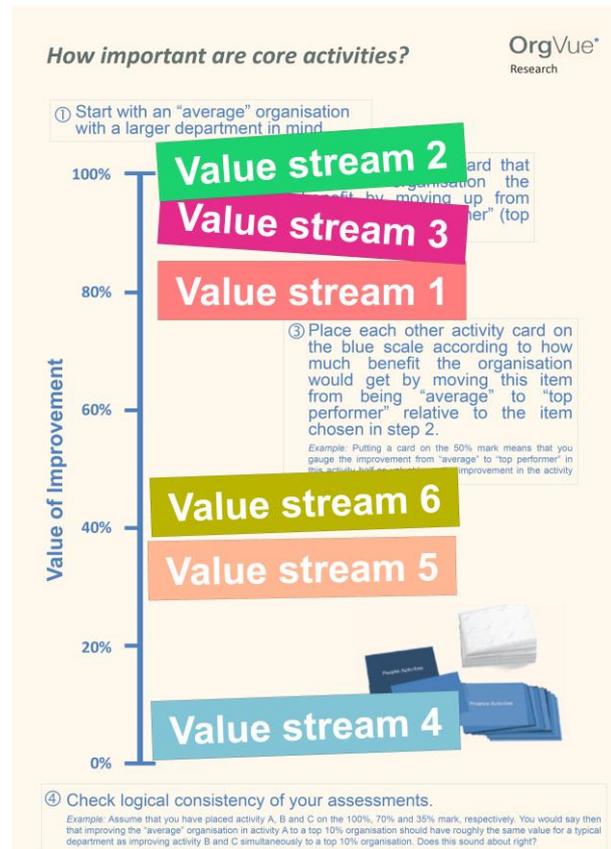

**Fig. 2.** Swing weighting with cards (example for 6 value streams).

Using classic swing weighting, the assessor $p$ is first asked to name their most important decision swing within process $i$ of value stream $h$, say $(h, i, \hat{j})$, (Bottomley & Doyle 2001). This swing receives a provisional weight $w^*_{hi\hat{j}p} = 100\%$. Doing so establishes a reference scale. For all other decisions $j$ in this process, the assessor is asked to estimate the value of swing $(h, i, j)$ relative to $(h, i, \hat{j})$. For instance, $w^*_{hij_1p} = 33\%$ and $w^*_{hij_2p} = 67\%$ would mean that improving $j_1$ and $j_2$ have, respectively, one-third and two-thirds of the value of improving decision $\hat{j}$. But improving decision $j_1$ and $j_2$ simultaneously should be judged as valuable as improving decision $\hat{j}$ by an assessor $p$ who provides consistent importance judgements (consistency check). The provisional weights $w^*_{hijp}$ of each participant are later normalised to $w_{hijp}$ such that $\sum_j w_{hijp} = 1 \ \forall (h, i)$. Afterwards, the geometric mean of the normalised weights $w_{hijp}$ from all participants is assigned to $w^*_{hij}$. The geometric mean is often used as simple consensus estimator by decision analysts in the absence of a real group discussion (e.g., Saaty 1990, Vaidya & Kumar 2006). Finally, all decision weights $w^*_{hij}$ are again normalised such that $\sum_j w_{hij} = 1 \ \forall (h, i)$.

---

[1] Instead of comparing the value of improving the decision-making capability for certain decisions, processes and value streams, some participants found it helpful to imagine how many things can realistically go wrong in an average organisation.

The process weights $w_{hi}$ and value-stream weights $w_h$ are also elicited by letting participants compare the average-to-top-performer swings in successful decision making for processes $i$ and value streams $h$, respectively. The overall weight $w_j$ of each single decision is the product $w_h w_{hi} w_{hij}$ with $h$ and $i$ referring, respectively, to the value stream and process to which decision $j$ belongs. Note that $\sum_j w_j = 1$.

To keep participants motivated to think deeply about the relative values of swings, they should be engaged in conversations about the reasons behind their judgements (Keeney 1992). Weighting consistency checks (French et al. 2009), as in the example given in the last-but-not-least paragraph, should also be carried out. To make the preference elicitation more interactive, participants can be asked to do the swing weighting with cards on A3 sheets with the instructions for the weighting method (see Fig. 2).

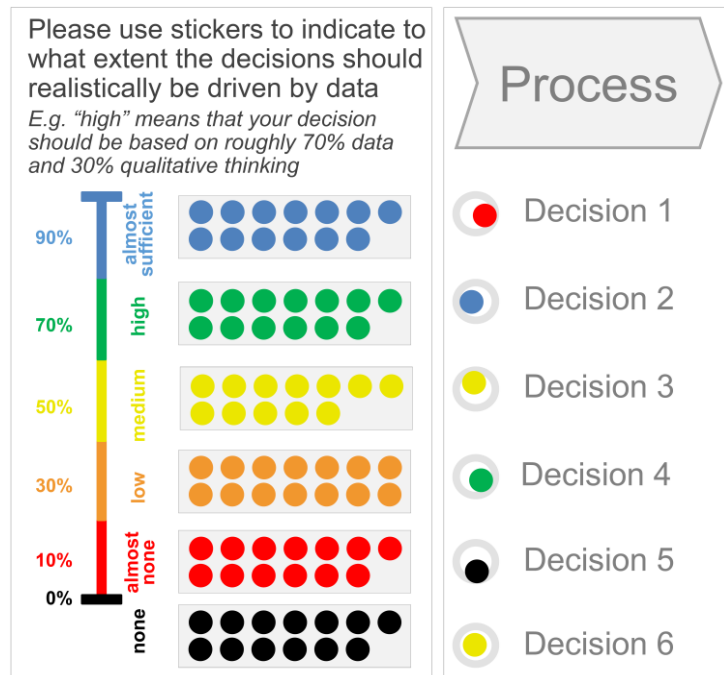

*Fig. 3.* Eliciting data support $d_j$ with a 0–1 scale and coloured stickers (example for a process with 6 decisions).

Instead of swing weighting, other methods such as AHP (Saaty 1990), MACEBTH (Bana e Costa & Vansnick 1994), SMART (Edwards 1977, Edwards & Barron 1994) and trade-off weighting (Keeney & Raiffa 1976) would certainly be equally appropriate from a technical perspective (Pöyhönen & Hämäläinen 2001). We chose swing weighting mainly because of its limited need for pairwise comparisons and its high technical transparency for non-experts. In line with research from Mau-Crimmins et al. (2005) and Schoemaker and Waid (1982), our hope is that the simplicity of this method will foster its acceptance and uptake by practitioners (see also Belton & Stewart 2002, Edwards et al. 1988, Keeney & von Winterfeldt 2007).

*3.3 Data support $d_j$ for making decisions*

Not all decisions can be informed equally well by business analytics. Long-range strategic decisions (e.g., 'How many different suppliers should we have?') are often harder to support with analytical reasoning than repetitive operational decisions (e.g., 'Do we need to intervene on our suppliers' delivery performance?'). Our proposed framework therefore incorporates the parameter data support $d_j$, which indicates to what extent decisions should realistically be driven by data. For each decision $j$, professionals with a strong analytical background are asked to provide estimates for $d_j$ on a 0–1 scale. For instance, $d_j = 0.4$ means that the decision should be based on 40% quantitative thinking and 60% qualitative thinking (e.g., experience, common sense, logic or observations). Analogous to the weights and again following the 'wisdom-of-the-crowd' thinking (Ariely et al. 2000, Surowiecki 2005), the geometric mean of the participants' data support estimates is later assigned to the decision.

In trials of the framework, participants judged it as difficult and tedious to assign precise numbers to $d_j$. To make obtaining parameters a more pleasant task, a symmetric Likert-type scale of measurement (Carifio & Rocco 2007, Likert 1932) with five response options is used instead: almost none ($d_j = 10\%$), low ($d_j = 30\%$), medium ($d_j = 50\%$), high ($d_j = 70\%$) and almost sufficient ($d_j = 90\%$). In addition, the response option 'no data support possible' ($d_j = 0\%$) is included for decisions that cannot be supported meaningfully by analytical reasoning. To make the survey a more pleasant activity, the participating professionals can be asked to put coloured circular stickers on each decision to indicate into which of the six available data-support classes they see it belonging. Fig. 3 shows an example of this assessment procedure.

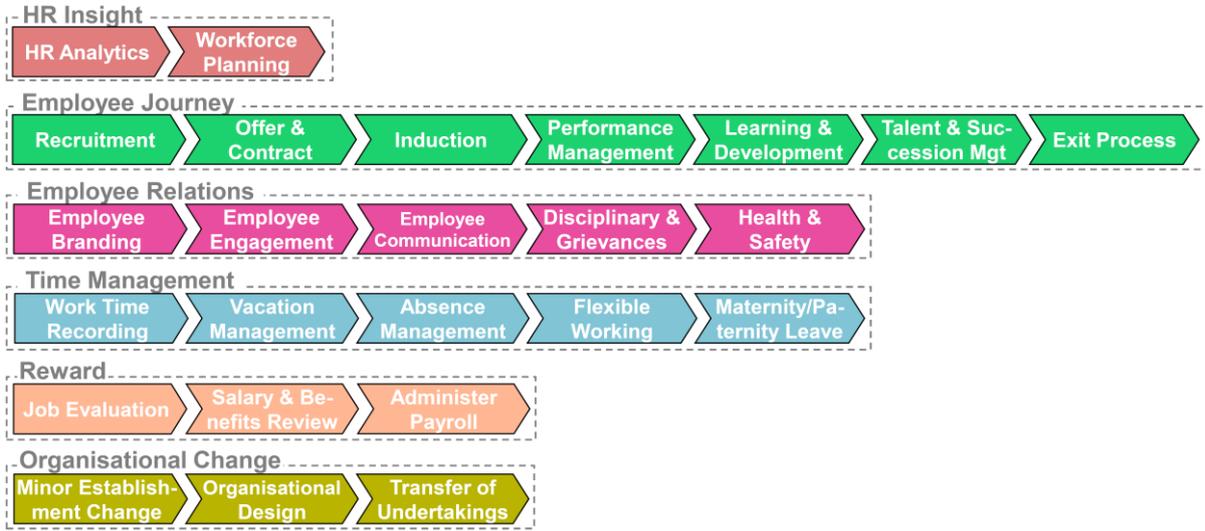

*Fig. 4.* HR process map.

*3.4 Priority index $I$ of data items*

By using analyses as an intermediate layer, the framework links data items to decisions. As an approximation, we gauge all data items as equally useful for informing the decisions to which they are linked. This approximation is necessary to avoid eliciting a weight parameter for each single analysis and each single data item; a drawback is that it makes the value of a data item within a decision dependent on the number of other data items linked to this decision. This simplification may lead to an over- or underestimation of the contribution of a data item to inform a particular decision. We also assume that even if the decision maker does not possess all data items required for a certain analysis, the responsible person may nevertheless be able to draw some conclusions from it. For example, educated guesses, industry surveys, simulations or break-even analyses might be used for the missing data items (e.g., Cascio & Boudreau 2011, Jackson 2007).

$\varphi_{jl} = 1$ if data item $l$ is linked (via at least one analysis) to decision $j$; otherwise 0. Also, let $n_j$ be the number of data items $l$ feeding into decision $j$. Furthermore, recall that the product $w_h w_{hi} w_{hij}$ is the overall weight of decision $j$ and $d_j$ adjusts this weight by how useful data support is for making this decision. The priority index for data item $l$ is defined as additive function

$$I(l) = \sum_h w_h \sum_i w_{hi} \sum_j w_{hij} d_j \frac{\varphi_{jl}}{n_j}.$$

$I(l)$ can be interpreted as an estimate of the overall relative importance of data item $l$ for improving an averagely run business function.

**4 Application to HR**

The development of the framework presented in this study was initiated and funded by a medium-sized consultancy specialising in business analytics. The consultancy wanted to provide clients of its HR software package with advice about what data they should store about their employees and HR activities. The project was conducted in the embedded research mode (Marshall et al. 2014, Wong 2009), an emerging trend in applied research where organisations employ and host academics to conduct rigorous in-house research that responds to the needs of the hosting organisation and the academic community.

## 4.1 Linking model

The process map captures the standard activities of an HR department as described in numerous handbooks (e.g., Armstrong & Taylor 2014, Bratton & Gold 2012, Mitchell & Gamlem 2012). While the grouping in chapters ('value streams') and subsections ('processes') varies between different books, a general consensus exists as to the main activities in HR. The process map depicted in Fig. 4 reflects the opinions of serval HR consultants and consultancy clients involved in the embedded research project on how to group these standard activities in a meaningful way.

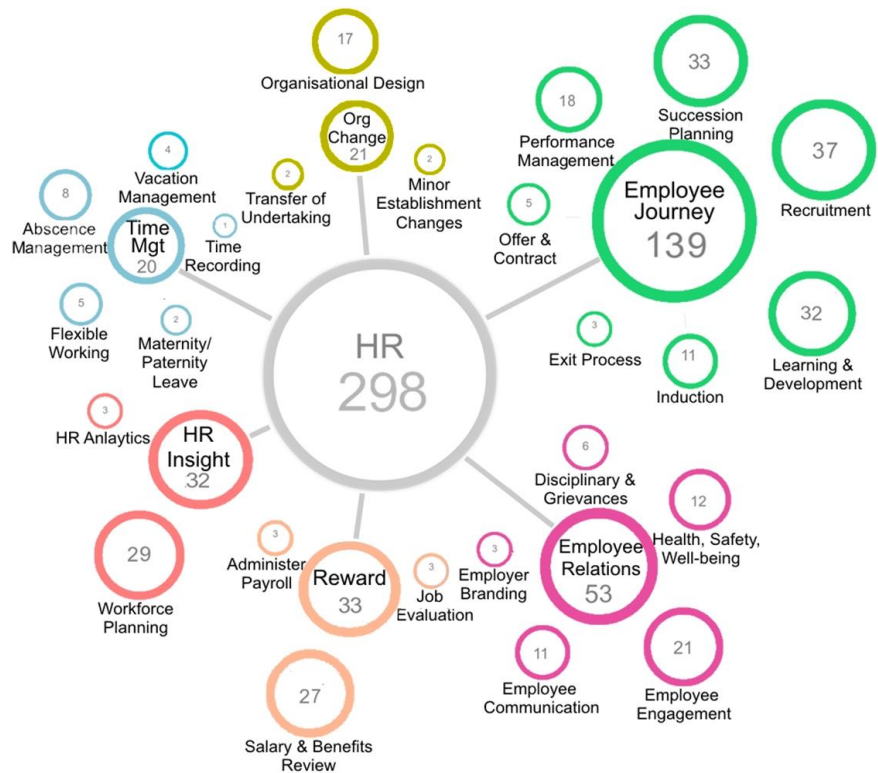

**Fig. 5.** Number of HR analyses by process.

Typical HR decisions include 'What actions should be taken on individual absence cases?' and 'How much spare capacity do we need for business continuity?' Through consulting HR handbooks and interviewing HR professionals, a total of 55 standard HR decisions were identified (see Appendix A). The list consists of operational decisions with a repetitive character and strategic one-off decisions.

Typical HR analyses include the correlation between performance score and compensation, prediction of number of employees expected to leave the company, time to fill vacancies, employee count and estimated monetary value of performance difference in role. We gathered 298 descriptive and predictive HR analyses and assigned them to the appropriate decisions. A total of 202 are selected from the academic and professional literature (e.g., Bassi et al. 2010, Becker et al. 2001, Boudreau 2010, Cascio & Boudreau 2011, Davenport et al. 2010, Fitz-enz 2010, Fitz-enz & Davison 2001, Infohrm 2010, Kavanagh & Thite 2009, SilkRoad 2012, Smith 2013) and the others from past consulting projects and suggestions provided by the HR analysts interviewed for this research. Fig. 5 counts the number of analyses belonging to each HR process. One may infer from this balloon chart that some HR processes (e.g., Recruitment and Workforce Planning) have received much attention by academics and HR analytics professionals in the last decades whereas others (e.g., Flexible Working and Organisational Design) have been less extensively addressed.

Examples of HR data items are competence assessment, date of birth, disciplinary date, email traffic between employees, having received particular HR communication message, job board usage data, participation in work-life balance programme, salary benchmarks, strategic importance of role, time to full productivity and user feedback on online application process. The set of HR analyses used in this case study requires 126 da-ta items grouped into 16 categories (see Table 1).

**Table 1**
Categories for HR data items.

| | |
|---|---|
| Absence | Hiring and induction |
| Application information | Motivation |
| Competences of employee | Other |
| Contract information | Personal details |
| Costs of employee | Performance and potentials |
| Disciplinary and grievances | Role information |
| Employee communication | Termination |
| Health & safety, well-being | Training |

*Table 2*
Participating HR professionals grouped by role and survey activity.

|  | Head of HR | Senior HR role | HR executive | HR consultant |
|---|---|---|---|---|
| *Weighting value streams* | 4 | 10 | 3 | 3 |
| *Weighting processes* | 3 | 7 | 2 | 3 |
| *Weighting decisions* | 2 | 4 | 1 | 1 |
| *Data support* | 2 | 6 | 0 | 1 |

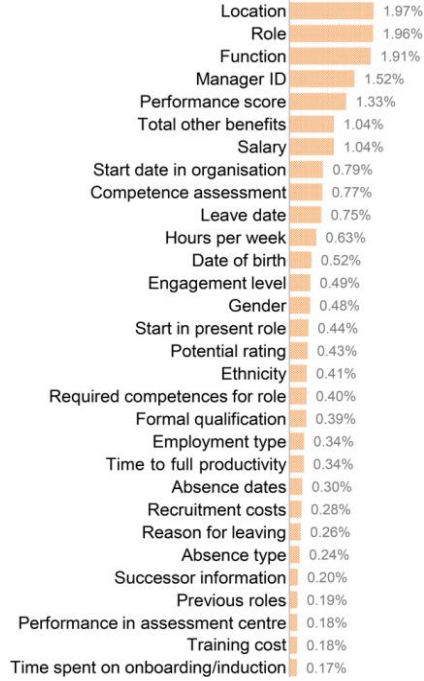

*Fig. 6.* Top 30 HR data items by priority index.

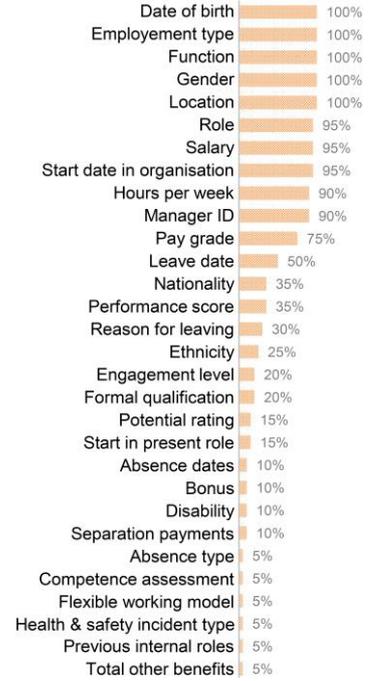

*Fig. 7.* Top 30 HR data items by percentage of availability in the 20 surveyed HR departments.

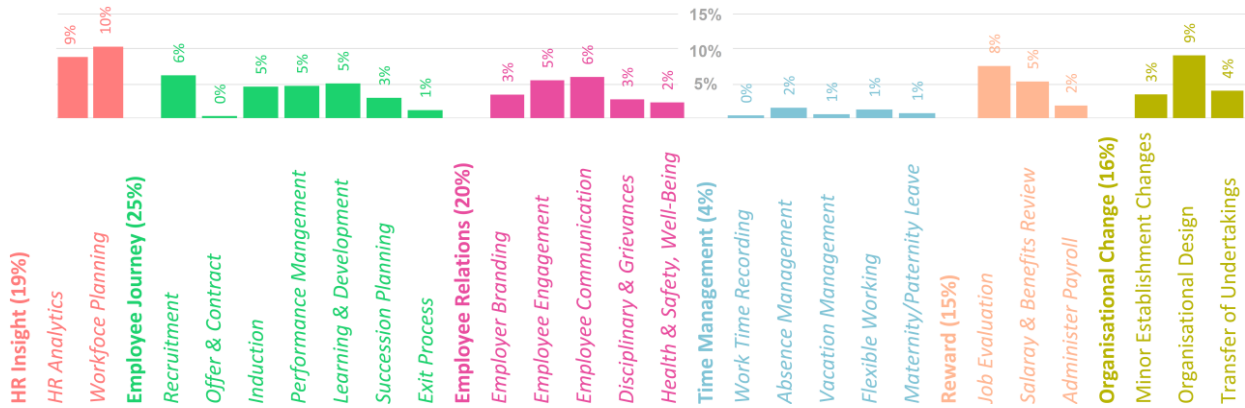

*Fig. 8.* Relative importance of improving each of the 25 HR processes.

### 4.2 Decision model

To elicit the parameters required by the decision-analysis component of the framework, 24 HR professionals from 15 organisations with a diverse mix of size, ownership and industry were interviewed (see Table 2). On average, the volunteers needed 10 minutes to weight the value streams, 25 minutes to weight the processes, 30 minutes to weight the decisions and 40 minutes to assess the data support for decisions. Most participants mentioned after the interview that they found physically 'playing' with the cards and circular stickers (as opposed to populating tables with numbers) fun and that they were not bored during the preference-elicitation session.

The overall HR process weights $w_i = w_h w_{hi}$ are depicted in Fig. 8. The decision weights and data support judgements are provided in Appendix A. Summing up the overall weighted data support (i.e., $\sum_h w_h \sum_i w_{hi} \sum_j w_{hij} d_j$), one may conclude that enhanced data collection and data use can contribute 36% of the effort required for an average HR department to become top performing in all its activities—a number judged as reasonable by the research participants.

The resulting top 30 HR data items according to the suggested framework are presented in Fig. 6. Since many HR analyses can be dimensionalised by location, role, function/department and manager ID, these four items inevitably received the highest scoring. The less common data items of time to full productivity and time spent on onboarding/induction are present in particular because they influence the organisation's turnover costs, which in turn is important for computing the return on investment for many HR activities.

*4.3 Comparison with current practice*

To compare the framework's suggestion on most-valued HR data items with current practices, 20 HR departments from 16 different countries were asked to share the data items they collected about their employees. The survey results are shown in Fig. 7.[1] A total of 23 items can be found in both top 30 lists, although in a quite different order. However, it also becomes clear that many of the surveyed companies collect hardly any HR data for business analytics beyond what is already available in their enterprise resource-planning software, despite the fact that 18 out of the surveyed 20 HR departments stated they already performed some form of HR analytics.

*4.4 Interpretation of results*

Reflecting on Fig. 6, the top 30 HR data items do not pose any difficulties to the technical capabilities of a conventional BI system. More general, slow-moving, linked 'people data' is at the heart of HR analytics (Slinger & Morrision 2014), i.e. employees being linked directly or indirectly to things such as roles, managers, competence requirements and recruitment costs. Therefore, HR datasets are in most cases not particularly demanding in terms of volume, velocity and variety to recall the original definition of big data (De Mauro et al. 2015). On the other hand, many HR departments will nevertheless find collecting those data items challenging for three reasons. First, to ensure that data items are measured in a valid, consistent and legal way, it may be necessary to invest resources in the development of new methods and policies for the collection of some data items (e.g., a validated staff survey and a toolkit to estimate the time to full productivity). Second, gathering some of the data items for all employees/roles in an organisation for the first time can be very time consuming (e.g., competence assessment and successor information), and the administrative time requirements for keeping these HR data items up to date are also not always negligible (Morrison 2015). Third, the central collection of some data items may require the purchase of and training sessions for new software packages/extensions to administer the data (e.g., training and absence management), which can be a substantial investment requiring a convincing business case.

Given that no external data sources (e.g., graduate statistics, job board data, reward benchmark data, social media presence) made it into the top 30 mirrors the current state of HR analytics: still inward-looking and just beginning to explore the opportunities of analytics (Deloitte 2014, Green 2014, Jacobs 2014). In fact, only very few of the 298 HR metrics which were gathered from the literature and HR professionals required the creation or acquisition of data items not typed in manually by staff. Applying the framework again in a few years' time may lead to a different top 30 list. When looking into the analytics literature of other business functions such as finance (e.g., Bragg 2014, Lee et al. 2009), marketing (e.g., Grigsby 2015, Venkatesan et al. 2014) and procurement (e.g., Feigin 2011, Pandit & Marmanis 2008), one would expect slightly more complex data items in terms of volume and velocity and to some extent also more variety in their top 30 lists.

It should be mentioned that constructing the linking model is not just about prioritising an already-existing list of potential data items. The linking model also offers a systematic way to gain a better awareness of the kind of HR analyses that are possible in different HR activity areas and of the kind of HR data items an organisation could collect.

---

[1] Note that the purely administrative data items like employee ID, tax number or email address are excluded from the list because they are usually not relevant for business analytics.

*4.5 Customisation of framework*

The 30 data items in Fig. 6 now form the standard data items together with some administrative data items like tax ID of a commercial HR software package. Using the new framework, the constancy where the present research was carried out has also begun to provide clients with more customised support on what HR data items they should gather about their employees. Clients prioritise HR data by swing weighting the value of improvement from 'where the business function is now' to 'where the business function wants to be' for each value stream, process and decision.

## 5 Conclusion

The present study contributes to the literature by proposing a framework based on multi-criteria decision analysis to estimate the value of data items for conducting business analytics. The novel framework can inform the discussion about data items that should be gathered routinely by a business function to analyse its activities. In our role as 'embedded researchers' at a consulting company, we identified a practical need for such a framework, and a systematic literature search showed that the problem of prioritising data items for business analytics has hitherto not been addressed in the academic literature. We validated the framework with an application to HR and produced a list of the top 30 data items for HR analytics.

Further research might focus on finding methods to overcome the limitations of our presented framework. First, the process of constructing the linking model is unsystematic in the sense that it does not prescribe how the literature should be consulted or how expert opinions should be enquired. System modelling languages such as the $i^*$ framework, the soft system methodology and Use Case have been previously employed to provide a richer perspective on the analytical needs of managers. The mixing of system modelling languages and decision analysis appears to be a promising approach to effectively explore and prioritise the data needs of a business function. Second, the framework does not consider the costs of gathering and maintaining different data items. Therefore, an extension of the framework, for instance, using portfolio decision analysis (Phillips & Bana e Costa 2007, Salo et al. 2011) to include the costs in the prioritisation exercise, would be desirable. Third, the framework works only for standard business analytics using conventional BI systems. Big data analytics, where the machine searches for correlations among a large set of potentially weak relevant data items, require a fundamentally different approach to prioritise the acquisition of datasets. Fourth, to date, the framework has only been applied to HR. Practical evidence regarding the use of the framework in other business functions such as finance, marketing, procurement and sales would be valuable.

The popularity of business analytic solutions is expected to continue to grow in industry and academia over the next years. A more systematic understanding of the benefits and costs associated with creating new data items for business analytics is likely to play a role in this development. We therefore believe there is merit in further exploring opportunities to apply and extend methods such as the one presented in this paper to quantify these benefits and costs.

**Acknowledgement:** I am grateful to Marianna Favero (Bocconi University) for leading the research on the HR process map and Giles Slinger (Concentra Consulting Ltd.) for reaching out to HR professionals and supporting the present research in numerous other ways. I also thank Ben Marshall, Robert Dyson, Rupert Morrison and three anonymous reviewers for advice on improving this paper.

## Appendix A: Complete list of HR decisions with parameters

| HR decisions | Weight $w_j$ | Data support $d_j$ |
|---|---|---|
| **HR Insight** | | |
| *HR Analytics* | | |
| What decisions do we want to support with HR analytics? | 3.15% | 0 |
| What reports do we want to provide to whom? | 2.25% | 0.1 |
| Which business outcomes correlate with HR metrics? | 3.49% | 0.1 |
| *Workforce Planning* | | |
| What mix of part-time, temporary and contract employees do we want? | 1.08% | 0.5 |
| Do we have the right diversity mix in terms of age, ethnicity, gender, disabilities and nationalities? | 1.44% | 0.7 |
| How much spare capacity do we need for business continuity? | 1.99% | 0.5 |
| How many people, of what skills do we need to hire in the next planning period in the different roles? | 2.61% | 0.5 |
| How do we address required change in headcount or skills (e.g., recruitment, training and takeover)? | 3.27% | 0.3 |

**Employee Journey**
*Recruitment*

| | | |
|---|---|---|
| Are we attracting the right new hires in sufficient quantity? | 1.79% | 0.5 |
| Are our pre-hire applicant assessments reliable? | 1.08% | 0.7 |
| Is our recruitment process efficient in terms of time and costs? | 0.67% | 0.7 |
| What channels to use to approach potential applicants (e.g., university, direct advertisement, agencies, internally)? | 1.40% | 0.5 |
| Whom shall we make an offer? | 1.28% | 0.5 |

*Offers & Contracts*

| | | |
|---|---|---|
| What terms should we offer to individual applicants? | 0.29% | 0.5 |

*Induction*

| | | |
|---|---|---|
| What should our induction/onboarding process be? | 2.29% | 0.3 |
| To whom should we offer which induction/onboarding? | 2.29% | 0.1 |

*Performance Management*

| | | |
|---|---|---|
| Which guidelines for measuring performance should we set (e.g., responsibilities, criteria, ...) , and is our measuring reliable? | 1.78% | 0.3 |
| Do we need to intervene on the performance of individual leaders? | 1.57% | 0.5 |
| What action should be taken about individual underperformers? | 1.33% | 0.3 |

*Learning & Development*

| | | |
|---|---|---|
| What methods should we use to identify talent, and is it reliable? | 1.76% | 0.5 |
| Is our overall learning and development approach helping us to achieve our HR strategy? | 1.50% | 0.5 |
| What informal learning processes should be encouraged in our organisation? | 0.72% | 0.1 |
| Which training courses should be offered? | 1.06% | 0.3 |

*Succession Management*

| | | |
|---|---|---|
| Is our succession management approach effective in identifying successors? | 0.87% | 0.7 |
| Which roles need nominated successors? | 0.90% | 0.5 |
| Who should we identify as a successor? | 0.65% | 0.5 |
| Who should be promoted? | 0.52% | 0.5 |

*Exit Process*

| | | |
|---|---|---|
| Who should manage the exit process? | 0.28% | 0.1 |
| What offer should we make potential leavers to stay? | 0.88% | 0.3 |

**Employee Relations**
*Employee Engagement*

| | | |
|---|---|---|
| Does our present employee engagement fit to our business strategy? | 1.71% | 0.5 |
| Do we need to intervene on staff turnover? | 1.71% | 0.7 |

*Employer Branding*

| | | |
|---|---|---|
| Does our employer branding fit to our business strategy? | 5.50% | 0.3 |

*Employee Communication & Consultation*

| | | |
|---|---|---|
| Do we need to make changes on our overall employee communication approach? | 2.77% | 0.3 |
| Do we need to make changes on our employee consultation approach? | 1.97% | 0.3 |
| What mix of employee communication methods should we use? | 1.25% | 0.3 |

*Disciplinary and Grievances*

| | | |
|---|---|---|
| What is our policy for detecting and dealing with non-compliant behaviour? | 0.83% | 0 |
| What are the root causes for disciplinary and grievance actions we deal with? | 1.05% | 0.5 |
| What action should be taken on individual disciplinary & grievances cases? | 0.84% | 0.1 |

*Health & Safety, Employee Well-Being*

| | | |
|---|---|---|
| Do we need to intervene on our health & safety policy? | 1.14% | 0.7 |
| Do we need to intervene on employee well-being? | 1.14% | 0.5 |

**Time Management**
*Time Recording*

| | | |
|---|---|---|
| Should we record the working time of our employees, and how? | 0.40% | 0.5 |

*Vacation Management*

| | | |
|---|---|---|
| How many days of vacation should we offer our employees? | 0.67% | 0.7 |
| How should we design the process and priorities for vacation requests? | 0.84% | 0.3 |

*Absence Management*

| | | |
|---|---|---|
| Do we need to intervene on root causes for absenteeism and presenteeism? | 0.33% | 0.7 |
| What action should be taken on individual absence cases? | 0.23% | 0.3 |

*Flexible Working*

| | | |
|---|---|---|
| What flexible working options suit our business strategy? | 1.24% | 0.3 |

*Maternity/Paternity Leave*

| | | |
|---|---|---|
| What should our policy and process for dealing with maternity/paternity leaves be? | 0.71% | 0.3 |

**Reward**
*Job Evaluation*

| | | |
|---|---|---|
| What should the organisation's job evaluation policy be? | 7.58% | 0.1 |

*Salary & Benefits Reviews*

| | | |
|---|---|---|
| How should we compensate our employees? | 5.33% | 0.5 |

*Administer Payroll*

| | | |
|---|---|---|
| Are service levels, costs and outcomes acceptable? | 1.83% | 0.7 |

**Organisational Change**
*Minor Establishment Changes*

| | | |
|---|---|---|
| Should we change the positions of individual employees? | 3.24% | 0.3 |

*Organisational Design*

| | | |
|---|---|---|
| How do we structure our people in the different function to deliver our business strategy? | 4.32% | 0.7 |
| What roles do we require for our business? | 4.07% | 0.5 |
| Should we carry out organisational design by ourselves or get external support? | 0.74% | 0.1 |

*Transfer of Undertakings*

| | | |
|---|---|---|
| Does our transfer of undertakings process work? | 3.68% | 0.3 |